\documentstyle[11pt,newpasp,twoside]{article}
\def\edcomment#1{\iffalse\marginpar{\raggedright\sl#1\/}\else\relax\fi}
\reversemarginpar

\begin{document}
\title{Multi-color observations of Young Star Clusters}
 \author{Peter Anders, Uta Fritze - v. Alvensleben}
\affil{Universit\"ats-Sternwarte G\"ottingen, Geismarlandstrasse 11, 37083 G\"ottingen, Germany}
\author{Richard de Grijs}
\affil{Department of Physics \& Astronomy, The University of Sheffield, Hicks Building, Hounsfield Road, Sheffield S3 7RH, UK}

\begin{abstract}
We present a new set of evolutionary synthesis models of our GALEV code, specifically developed to include the gaseous emission of presently forming star clusters, in combination with an advanced tool to compare large model grids with multi-color broad-band observations of YSC systems. Tests and first applications are presented.
\end{abstract}

\section{Models \& Applications}
We have further refined the G\"ottingen evolutionary synthesis code GALEV by including the effect of gaseous emission. The emission contributes significantly to the integrated light of stellar populations younger than $3 \times 10^7$ yr (Anders \& Fritze - v. Alvensleben 2003). The updated models are available from {\bf http://www.uni-sw.gwdg.de/$\sim$galev/panders/}.\\
The tool to compare large model grids with multi-color broad-band observations was tested extensively using artificial clusters (Anders et al. 2003a) and broad-band observations of star clusters in NGC 3310 (de Grijs et al. 2003a).\\
Here we focus on the dwarf starburst galaxy NGC 1569, in which we detect and analyse a sample of star clusters significantly larger than done before. Our derived cluster properties are consistent with literature values. We find a surprising dependence of the (cluster) mass function on cluster ages (Anders et al. 2003b).


\section{References}
Anders, P., Fritze - v. Alvensleben, U. 2003, \aap, 401, 1063\\
Anders, P., Bissantz, N., Fritze - v. Alvensleben, U., de Grijs, R. 2003a,\\ \mnras, {\sl submitted}\\
Anders, P., de Grijs, R., Fritze - v. Alvensleben, U., Bissantz, N. 2003b,\\ \mnras, {\sl submitted}\\
de Grijs, R., Fritze - v. Alvensleben, U., Anders, P., Gallagher, J. S., Bastian, N., Taylor, V. A., Windhorst, R. A. 2003a, \mnras, 342, 259\\
\end{document}